\documentclass[%
reprint,
superscriptaddress,
nofootinbib,
 amsmath,amssymb,
 aps,
 prc,
]{revtex4-2}

\usepackage{graphicx}% Include figure files
\usepackage{dcolumn}% Align table columns on decimal point
\usepackage{bm}% bold math
\usepackage{hyperref}% add hypertext capabilities
\usepackage{float}
\usepackage{multirow}
\usepackage{braket}
\usepackage{placeins}

\begin{document}

\preprint{APS/PRL}

\title{Nuclear density dependence of polarization transfer\\ in quasi-elastic ${\rm A}(\vec{e},e' \vec{p})$ reactions}% Force line breaks with \\

\author{T.~Kolar}
 \email{tkolar@mail.tau.ac.il}
 \affiliation{School of Physics and Astronomy, Tel Aviv University, Tel Aviv 69978, Israel.}
\author{W.~Cosyn}
 \affiliation{Florida International University, Miami, FL 33199, USA.}
\author{C.~Giusti}
  \affiliation{INFN, Sezione di Pavia, via A.~Bassi 6, I-27100 Pavia, Italy.}
\author{P.~Achenbach}
  \altaffiliation[Present address:]{Thomas Jefferson National Accelerator Facility, Newport News, VA 23606, USA.}
  \affiliation{Institut f\"ur Kernphysik, Johannes Gutenberg-Universit\"at, 55099 Mainz, Germany.}
\author{A.~Ashkenazi}
  \affiliation{School of Physics and Astronomy, Tel Aviv University, Tel Aviv 69978, Israel.}
\author{R.~B\"ohm}
  \affiliation{Institut f\"ur Kernphysik, Johannes Gutenberg-Universit\"at, 55099 Mainz, Germany.}
\author{D.~Bosnar} 
  \affiliation{Department of Physics, Faculty of Science, University of Zagreb, HR-10000 Zagreb, Croatia}
\author{T.~Brecelj}
  \affiliation{Jo\v{z}ef Stefan Institute, 1000 Ljubljana, Slovenia.}
\author{M.~Christmann}
  \affiliation{Institut f\"ur Kernphysik, Johannes Gutenberg-Universit\"at, 55099 Mainz, Germany.}
\author{E.O.~Cohen}
  \altaffiliation[Also at ]{Soreq NRC, Yavne 81800, Israel.}
  \affiliation{School of Physics and Astronomy, Tel Aviv University, Tel Aviv 69978, Israel.}
\author{M.O.~Distler}
  \affiliation{Institut f\"ur Kernphysik, Johannes Gutenberg-Universit\"at, 55099 Mainz, Germany.}
\author{L.~Doria}
  \affiliation{Institut f\"ur Kernphysik, Johannes Gutenberg-Universit\"at, 55099 Mainz, Germany.}
\author{P.~Eckert}
  \affiliation{Institut f\"ur Kernphysik, Johannes Gutenberg-Universit\"at, 55099 Mainz, Germany.}
\author{A.~Esser}
  \affiliation{Institut f\"ur Kernphysik, Johannes Gutenberg-Universit\"at, 55099 Mainz, Germany.}
\author{J.~Geimer}
  \affiliation{Institut f\"ur Kernphysik, Johannes Gutenberg-Universit\"at, 55099 Mainz, Germany.}
\author{R.~Gilman}
  \affiliation{Rutgers, The State University of New Jersey, Piscataway, NJ 08855, USA}
\author{P.~G\"ulker}
  \affiliation{Institut f\"ur Kernphysik, Johannes Gutenberg-Universit\"at, 55099 Mainz, Germany.}
\author{M.~Hoek}
  \affiliation{Institut f\"ur Kernphysik, Johannes Gutenberg-Universit\"at, 55099 Mainz, Germany.}
\author{D.~Izraeli}
  \affiliation{School of Physics and Astronomy, Tel Aviv University, Tel Aviv 69978, Israel.}
\author{S.~Kegel}
  \affiliation{Institut f\"ur Kernphysik, Johannes Gutenberg-Universit\"at, 55099 Mainz, Germany.}
\author{P.~Klag}
  \affiliation{Institut f\"ur Kernphysik, Johannes Gutenberg-Universit\"at, 55099 Mainz, Germany.}
\author{I.~Korover}
  \altaffiliation[Also at]{Department of Physics, NRCN, P.O. Box 9001, Beer-Sheva 84190, Israel}
  \affiliation{School of Physics and Astronomy, Tel Aviv University, Tel Aviv 69978, Israel.}
\author{J.~Lichtenstadt}
  \affiliation{School of Physics and Astronomy, Tel Aviv University, Tel Aviv 69978, Israel.}
\author{M.~Littich}
  \affiliation{Institut f\"ur Kernphysik, Johannes Gutenberg-Universit\"at, 55099 Mainz, Germany.}
\author{T.~Manoussos}
  \affiliation{Institut f\"ur Kernphysik, Johannes Gutenberg-Universit\"at, 55099 Mainz, Germany.}
\author{I.~Mardor}
  \affiliation{School of Physics and Astronomy, Tel Aviv University, Tel Aviv 69978, Israel.}
  \affiliation{Soreq NRC, Yavne 81800, Israel.}
\author{D.~Markus}
  \affiliation{Institut f\"ur Kernphysik, Johannes Gutenberg-Universit\"at, 55099 Mainz, Germany.}
\author{H.~Merkel}
  \affiliation{Institut f\"ur Kernphysik, Johannes Gutenberg-Universit\"at, 55099 Mainz, Germany.}
\author{M.~Mihovilovi\v{c} }
  \affiliation{Faculty of Mathematics and Physics, University of Ljubljana, 1000 Ljubljana, Slovenia.}
  \affiliation{Jo\v{z}ef Stefan Institute, 1000 Ljubljana, Slovenia.}
  \affiliation{Institut f\"ur Kernphysik, Johannes Gutenberg-Universit\"at, 55099 Mainz, Germany.}
\author{J.~M\"uller}
  \affiliation{Institut f\"ur Kernphysik, Johannes Gutenberg-Universit\"at, 55099 Mainz, Germany.}
\author{U.~M\"uller}
  \affiliation{Institut f\"ur Kernphysik, Johannes Gutenberg-Universit\"at, 55099 Mainz, Germany.}
\author{M.~Olivenboim}
  \affiliation{School of Physics and Astronomy, Tel Aviv University, Tel Aviv 69978, Israel.}
\author{J.~P\"{a}tschke}
  \affiliation{Institut f\"ur Kernphysik, Johannes Gutenberg-Universit\"at, 55099 Mainz, Germany.}
\author{S.J.~Paul}
  \affiliation{Department of Physics and Astronomy, University of California, Riverside, CA 92521, USA.}
\author{E.~Piasetzky}
  \affiliation{School of Physics and Astronomy, Tel Aviv University, Tel Aviv 69978, Israel.}
\author{S.~Plura}
  \affiliation{Institut f\"ur Kernphysik, Johannes Gutenberg-Universit\"at, 55099 Mainz, Germany.}
\author{J.~Pochodzalla}
  \affiliation{Institut f\"ur Kernphysik, Johannes Gutenberg-Universit\"at, 55099 Mainz, Germany.}
\author{M.~Po\v{z}un}
  \affiliation{Jo\v{z}ef Stefan Institute, 1000 Ljubljana, Slovenia.}
\author{G.~Ron}
  \affiliation{Racah Institute of Physics, Hebrew University of Jerusalem, Jerusalem 91904, Israel.}
\author{B.S.~Schlimme}
  \affiliation{Institut f\"ur Kernphysik, Johannes Gutenberg-Universit\"at, 55099 Mainz, Germany.}
\author{M.~Schoth}
  \affiliation{Institut f\"ur Kernphysik, Johannes Gutenberg-Universit\"at, 55099 Mainz, Germany.}
\author{F.~Schulz}
  \affiliation{Institut f\"ur Kernphysik, Johannes Gutenberg-Universit\"at, 55099 Mainz, Germany.}
\author{C.~Sfienti}
  \affiliation{Institut f\"ur Kernphysik, Johannes Gutenberg-Universit\"at, 55099 Mainz, Germany.}
\author{S.~\v{S}irca}
  \affiliation{Faculty of Mathematics and Physics, University of Ljubljana, 1000 Ljubljana, Slovenia.}
  \affiliation{Jo\v{z}ef Stefan Institute, 1000 Ljubljana, Slovenia.}
\author{R.~Spreckels}   %OK
  \affiliation{Institut f\"ur Kernphysik, Johannes Gutenberg-Universit\"at, 55099 Mainz, Germany.}
\author{S.~\v{S}tajner }
  \affiliation{Jo\v{z}ef Stefan Institute, 1000 Ljubljana, Slovenia.}
\author{S.~Stengel}
  \affiliation{Institut f\"ur Kernphysik, Johannes Gutenberg-Universit\"at, 55099 Mainz, Germany.}
\author{E.~Stephan}
  \affiliation{Institute of Physics, University of Silesia in Katowice, 41-500 Chorz\'ow, Poland.}
\author{Y.~St\"ottinger}   %Kohl OK
  \affiliation{Institut f\"ur Kernphysik, Johannes Gutenberg-Universit\"at, 55099 Mainz, Germany.}
\author{S.~Strauch}
  \affiliation{University of South Carolina, Columbia, South Carolina 29208, USA.}
\author{C.~Szyszka}
  \affiliation{Institut f\"ur Kernphysik, Johannes Gutenberg-Universit\"at, 55099 Mainz, Germany.}
\author{M.~Thiel}
  \affiliation{Institut f\"ur Kernphysik, Johannes Gutenberg-Universit\"at, 55099 Mainz, Germany.}
\author{A.~Weber} %OK
  \affiliation{Institut f\"ur Kernphysik, Johannes Gutenberg-Universit\"at, 55099 Mainz, Germany.}
\author{A.~Wilczek}
  \affiliation{Institute of Physics, University of Silesia in Katowice, 41-500 Chorz\'ow, Poland.}
\author{I.~Yaron}
  \affiliation{School of Physics and Astronomy, Tel Aviv University, Tel Aviv 69978, Israel.}
\collaboration{\textbf{A1 Collaboration}}
\noaffiliation

\date{\today}% It is always \today, today,

\begin{abstract}
The ratio of the transverse and longitudinal components of polarization transfer to protons in the quasi-elastic $(\vec{e}, e^{\prime} \vec{p}\,)$ reaction, $P^{\prime}_x/P^{\prime}_z$, is sensitive to the proton's electromagnetic form factor ratio, $G_E/G_M$. To explore density-dependent in-medium modifications, a comparison of polarization transfer ratios involving protons from distinct nuclear shells, each with different local nuclear densities, has been proposed. In this study, we present such comparisons between four shells, $1s_{1/2}$, $1p_{3/2}$ in $^{12}\mathrm{C}$ and $1d_{3/2}$, $2s_{1/2}$ in $^{40}\mathrm{Ca}$. In an effort to account for other many-body effects that may differ between shells, we use a state-of-the-art relativistic distorted-wave impulse-approximation (RDWIA) calculation and present the \textit{double ratios}, $(P^{\prime}_x/P^{\prime}_z)_{\rm Data}/(P^{\prime}_x/P^{\prime}_z)_{\rm RDWIA}$ as well as the \textit{super ratios}, $\left[(P^{\prime}_x/P^{\prime}_z)_{\rm A}/(P^{\prime}_x/P^{\prime}_z)_{\rm B}\right]_{\rm Data}/\left[(P^{\prime}_x/P^{\prime}_z)_{\rm A}/(P^{\prime}_x/P^{\prime}_z)_{\rm B}\right]_{\rm RDWIA}$, for chosen shells A and B, as a function of effective local nuclear densities. We find that double ratios for individual shells show a dependence on the probed effective nuclear densities. Studying the super ratios, we observed a systematic variation between pairs of higher- and lower-density shells.
\end{abstract}

\maketitle

%---------------------------------------------
%----------------  SECTION -------------------
%---------------------------------------------
\section{Introduction}
Polarization transfer to a proton bound in a nucleus has been suggested as a tool to observe in-medium modifications in the bound proton structure~\cite{Kelly:1996hd}. It is a part of a wider effort to understand the role of quarks and gluons in nuclei~\cite{Cloët_2019}. Some calculations introduce in-medium modifications and suggest nuclear-density-dependent changes of the bound nucleon electromagnetic (EM) form factors (FFs)~\cite{Smith:2004dn,Lu:1997mu, Ron:2012cp}. We report on the first systematic search for nuclear density-dependent effects in the quasi-elastic $A(\vec{e},e^{\prime}\vec{p})$ reaction, which is sensitive to EM FFs. 

For a free proton, the ratio of the transverse and longitudinal polarization-transfer components, $P^{\prime}_x/P^{\prime}_z$, in polarized elastic electron scattering, under the one-photon exchange approximation, is proportional to the proton EM form-factor ratio, $G_E/G_M$. Similarly, a polarization transfer in the quasi-elastic $\mathrm{A}(\vec{e}, e^{\prime} \vec{p}\,)$ reaction is sensitive to the \textit{effective} EM FFs which are related to the charge and magnetization distributions of the bound proton. However, quasi-elastic reactions are subject to other many-body effects, such as final state interactions, isobar configurations, and meson-exchange currents, which need to be well understood in order to isolate possible deviations due to modifications in the proton structure. 

It has been suggested to study the \textit{double ratios}~\cite{Ron:2012cp}, $(P^{\prime}_x/P^{\prime}_z)_{\rm Data}/(P^{\prime}_x/P^{\prime}_z)_{\rm Calc}$, thus dividing out the nuclear many-body effects included in the calculation. Furthermore, pairs of different shells (shell A and B), characterized by different nuclear densities, can be compared by the \textit{super ratios}, $\left[(P^{\prime}_x/P^{\prime}_z)_{\rm A}/(P^{\prime}_x/P^{\prime}_z)_{\rm B}\right]_{\rm Data}/\left[(P^{\prime}_x/P^{\prime}_z)_{\rm A}/(P^{\prime}_x/P^{\prime}_z)_{\rm B}\right]_{\rm Calc}$, looking for density-dependent medium modifications. The super ratios account for those differences in many-body effects that are included in the model, and reduce the sensitivity to systematic discrepancies common to calculations for different shells. 

Density-dependent modifications are expected to be at the level of a few percent~\cite{Ron:2012cp}, which requires high statistical accuracy. $^{12}\mathrm{C}$ was suggested as a good nucleus for such studies since the effective local nuclear density experienced by the protons bound in the $1s_{1/2}$ shell is about twice the density for those in the $1p_{3/2}$ shell. Comparison of polarization ratios for protons from $1s_{1/2}$ and $1p_{1/2}$ shells in $^{12}\mathrm{C}$ has shown that the $p$ to $s$ double ratio was $1.15\pm 0.03$~\cite{ceepTim}. Relativistic distorted-wave impulse-approximation (RDWIA) calculations only partially accounted for this deviation from unity, while the relativistic plane-wave impulse-approximation (RPWIA) did not predict this difference. However, once protons were compared at the same virtuality, the results became consistent with unity, $1.05\pm 0.05$, and so did RPWIA and RDWIA calculations~\cite{ceepTim}. We note that as seen in Fig.~2 of Ref.~\cite{ceepTim}, in the virtuality overlap region, we are effectively comparing the low-$p_{\rm miss}$ region of $1s_{1/2}$ to the high-$p_{\rm miss}$ region of $1p_{3/2}$. This reduces the difference in the probed effective densities between the two shells (see Fig.~\ref{fig:pmiss_and_density}). %Thus no attempt has been made to study further the \textit{super ratios}.

In this work we re-analyze recent data obtained for protons from the $1d_{3/2}$ and $2s_{1/2}$ shells in $^{40}\mathrm{Ca}$~\cite{caeepTim} and $1s_{1/2}$ and $1p_{1/2}$ shells in $^{12}\mathrm{C}$ under two different kinematic settings~\cite{ceepComp, ceepTim}. We compare the experimentally obtained ratios of polarization transfer components with corresponding RDWIA calculations, and preform the first systematic search for nuclear density-dependent effects in $A(\vec{e},e^{\prime}\vec{p})$.
%---------------------------------------------
%----------------  SECTION -------------------
%---------------------------------------------
\section{Polarization transfer data}
The polarization transfer components from the quasi-elastic $\mathrm{A}(\vec{e}, e^{\prime} \vec{p}\,)$ reaction on $^{12}\mathrm{C}$ and $^{40}\mathrm{Ca}$ were measured at the three-spectrometer facility of the A1 Collaboration at the Mainz Microtron (MAMI), using the $600\,\mathrm{MeV}$ polarized continuous-wave electron beam. The scattered electrons and the knocked-out protons were detected in coincidence using two magnetic spectrometers. The polarization components were measured with a polarimeter located near the focal plane of the proton spectrometer. These measurements were reported in Refs.~\cite{ceepComp,ceepTim,caeepTim}, and their kinematic parameters are summarized in Table~\ref{tab:kinematics}. We follow~\cite{ceepTim} and define the scalar proton missing momentum $p_{\rm miss}\equiv\pm|\vec{p}_{\rm miss}|=\pm |\vec{q}-\vec{p}^{\prime}|$, where $\vec{q}$ and $\vec{p}$ are the momentum transfer and the outgoing proton momentum, respectively. The sign is taken to be positive (negative) if the longitudinal component of $\vec{p}_{\rm miss}$ is parallel (anti-parallel) to $\vec{q}$.

The $^{12}\mathrm{C}$ data sets cover two ranges in $p_{\rm miss}$: the low-$p_{\rm miss}$ setting is centered around $p_{\rm miss}=0\,\mathrm{MeV/c}$ and extends to $p_{\rm miss}=\pm 140\,\mathrm{MeV/c}$, while the high-$p_{\rm miss}$ data range from $-260\,\mathrm{MeV/c}$ to $-100\,\mathrm{MeV/c}$. The data obtained for $^{40}\mathrm{Ca}$ span the range of $-200\,\mathrm{MeV/c}<p_{\rm miss}< -20\,\mathrm{MeV/c}$. The shell from which the proton was ejected was determined from the missing energy of each event. The measured missing momentum spectrum for each data set is shown in the left column of Fig.~\ref{fig:pmiss_and_density}.

\begin{table}
\caption{%\doublespacing
Kinematic settings of the ${\rm A}(\vec{e},e^{\prime}\vec{p}\,)$ measurements considered in this work. Following the beam energy, $E_{\rm beam}$, and the square of the transferred four-momentum, $Q^2$, we list the missing momentum ranges covered and its average for each considered shell. For $p_e$ and $\theta_e$ ($p_p$ and $\theta_p$) we denote the scattered electron (knocked-out proton) central momentum and angle settings, respectively.  
}
\begin{center}
\begin{tabular}{l@{}l@{}rrr}
\hline\hline\\[-8pt]
\multicolumn{2}{l}{\makebox[0pt][l]{Kinematic setting}} & $^{12}\mathrm{C}$-low & $^{12}\mathrm{C}$-high & $^{40}\mathrm{Ca}$  \\[2pt]
\hline\\[-8pt]
$E_{\rm beam}$			& [MeV]                   & 600				    & 600			        & 600			\\[3pt]
$Q^2$          			& [$({\rm GeV}\!/\!c)^2$] & 0.40         		& 0.18 			        & 0.25         		\\[3pt]
$p_{\rm miss}$ 			& [MeV$\!/\!c$]           & $[-150,0]$   		& $[-260,-100]$		    & $[-210,-17]$ 		\\[3pt]
$\braket{p_{\rm miss}}$	& [MeV$\!/\!c$]           & $1p_{3/2}$: $-$82   & $1p_{3/2}$: $-$171 	& $1d_{3/2}$: $-$123    \\[2pt]
               			&                         & $1s_{1/2}$: $-$60   & $1s_{1/2}$: $-$161 	& $2s_{1/2}$: $\;\,-$72 \\[3pt]
$p_e$          			& [MeV$\!/\!c$]           & 384          		& 368	      		    & 396          		\\[3pt]
$\theta_e$     			& [deg]                   & 82.4         		& 52.9	      		    & 61.8         		\\[3pt]
$p_p$          			& [MeV$\!/\!c$]           & 668          		& 665			        & 630          		\\[3pt]
$\theta_{p}$   			& [deg]                   & 34.7         		& 37.8			        & 40.2         		\\[3pt]
\hline\hline
\end{tabular}
\end{center}
\label{tab:kinematics}
\end{table}

%---------------------------------------------
%----------------  SECTION -------------------
%---------------------------------------------
\section{Calculations}
%----------------  SUBSECTION ----------------
\subsection{Polarization transfer}\label{subsec:PolTransCalc}
The polarization transfer for each data set was calculated with the RDWIA model of Ref.~\cite{Meucci:2001qc} using free-proton EM form factors. Our previous analyses have shown that the calculated results are in good agreement with the measured polarization transfer data~\cite{ceepComp,ceepTim,caeepTim}. 

The calculations were performed on an event-by-event basis. Using each event's kinematics parameters ensures a full match of the calculation to the experimental kinematics acceptance. It also allowed us to extract per-bin averages of the calculated observables. The original RDWIA program~\cite{Meucci:2001qc} was modified to include all 18 hadronic structure functions for the $\mathrm{A}(\vec{e}, e^{\prime} \vec{p})$ reaction in the Born approximation~\cite{Boffi:1996ikg}. The RDWIA calculations use the global democratic relativistic optical potential~\cite{Cooper:2009}, relativistic bound-state wave functions obtained with the NL-SH parametrization~\cite{SHARMA1993377}, and free-proton EM FFs using the Bernauer parametrization~\cite{Bernauer}. 

These calculations were compared to the measured polarization transfer data. More details are available in Refs.~\cite{ceepComp,ceepTim,HelictyAsymmC12Kolar,InducedC12,caeepTim}. The impact of deficiencies in the calculations can be further reduced by studying the ratios, $P^{\prime}_x/P^{\prime}_z$, rather than the individual components, $P^{\prime}_x$ and $P^{\prime}_z$. The double ratio, $(P^{\prime}_x/P^{\prime}_z)_{\rm Data}/(P^{\prime}_x/P^{\prime}_z)_{\rm RDWIA}$, factors out the many-body effects in the quasi-elastic process which are accounted for in the calculation. We note that in parallel/antiparallel kinematics the calculations depend linearly on the proton EM FFs ratio~\cite{caeepTim}.

\begin{figure}[ht]
\includegraphics[width=\columnwidth]{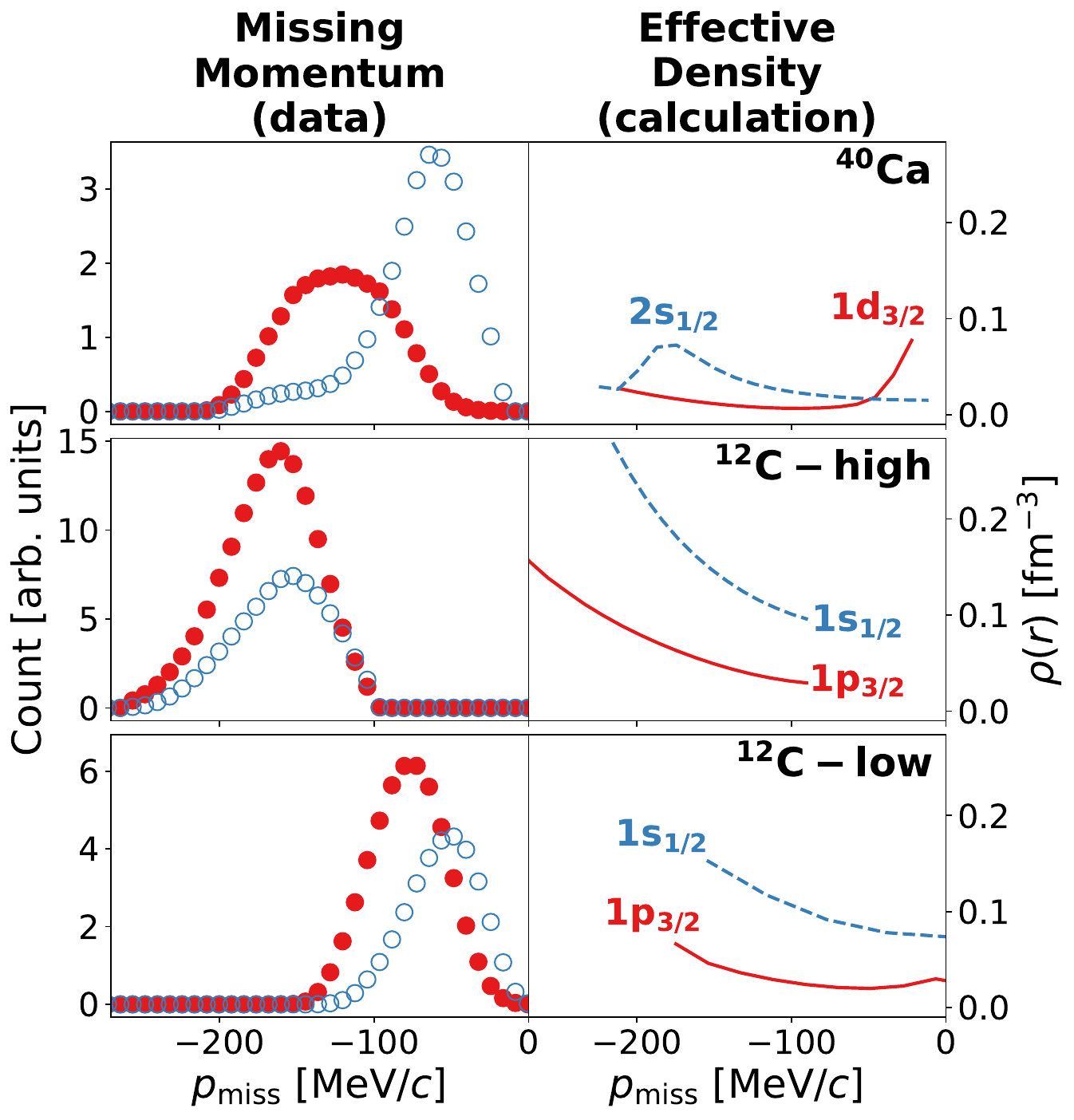}
\caption{Measured missing momentum distribution (left) and effective local nuclear densities (right) calculated in RDWIA for the three kinematic settings from Table~\ref{tab:kinematics}.
}
\label{fig:pmiss_and_density}
\end{figure}
%----------------  SUBSECTION ----------------
\subsection{Effective nuclear densities}\label{subsec:AvgNuclDensitiesCalc}
The effective local densities for protons removed from different shells in $^{12}\mathrm{C}$ and $^{40}\mathrm{Ca}$ have been obtained by following the procedure described in Refs.~\cite{Ryckebusch:2011yy, Ron:2012cp} but in RDWIA using the same model discussed above. In RDWIA the $(e,e^{\prime}p)$ cross section is not factorized and the distorted momentum distribution from~\cite{Ryckebusch:2011yy} corresponds to the so-called reduced cross section~\cite{Meucci:2001qc,Boffi:1996ikg}. The reduced cross section is obtained by dividing the cross section by a kinematical factor and the elementary off-shell electron-proton scattering cross-section for which we used the cc2 prescription of de Forest~\cite{DEFOREST1983232}. This way we obtain a spectral-function-like dependence solely on $E_{\rm miss}$ and $p_{\rm miss}$, but with included FSI and other many-body effects. We note that in the non-relativistic PWIA, the reduced cross section gives the momentum distribution of the bound proton wave function and, in a factorized DWIA, the so-called distorted momentum distribution.

The calculated effective densities are shown in Fig.~\ref{fig:pmiss_and_density} (right). We note the large differences between the effective densities of the $s$ and $p$ shells in $^{12}\mathrm{C}$. Similar differences were predicted in Ref.~\cite{Ron:2012cp} and suggested for studies of in-medium effects on the bound proton structure. The effective local densities in the $s$ and $d$ shells of $^{40}\mathrm{Ca}$ are similar, thus differences between these shells are expected to be smaller. A comparison of a shell in $^{12}\mathrm{C}$ with a shell in $^{40}\mathrm{Ca}$ can serve as a cross-check to density-dependent modifications, but it is more susceptible to systematic uncertainties and deficiencies in the calculations.
%---------------------------------------------
%----------------  SECTION -------------------
%---------------------------------------------
\section{Density dependence of the polarization transfer}
%----------------  SUBSECTION ----------------
\subsection{Single-shell comparison with RDWIA}
\begin{figure}[b]
\includegraphics[width=\columnwidth]{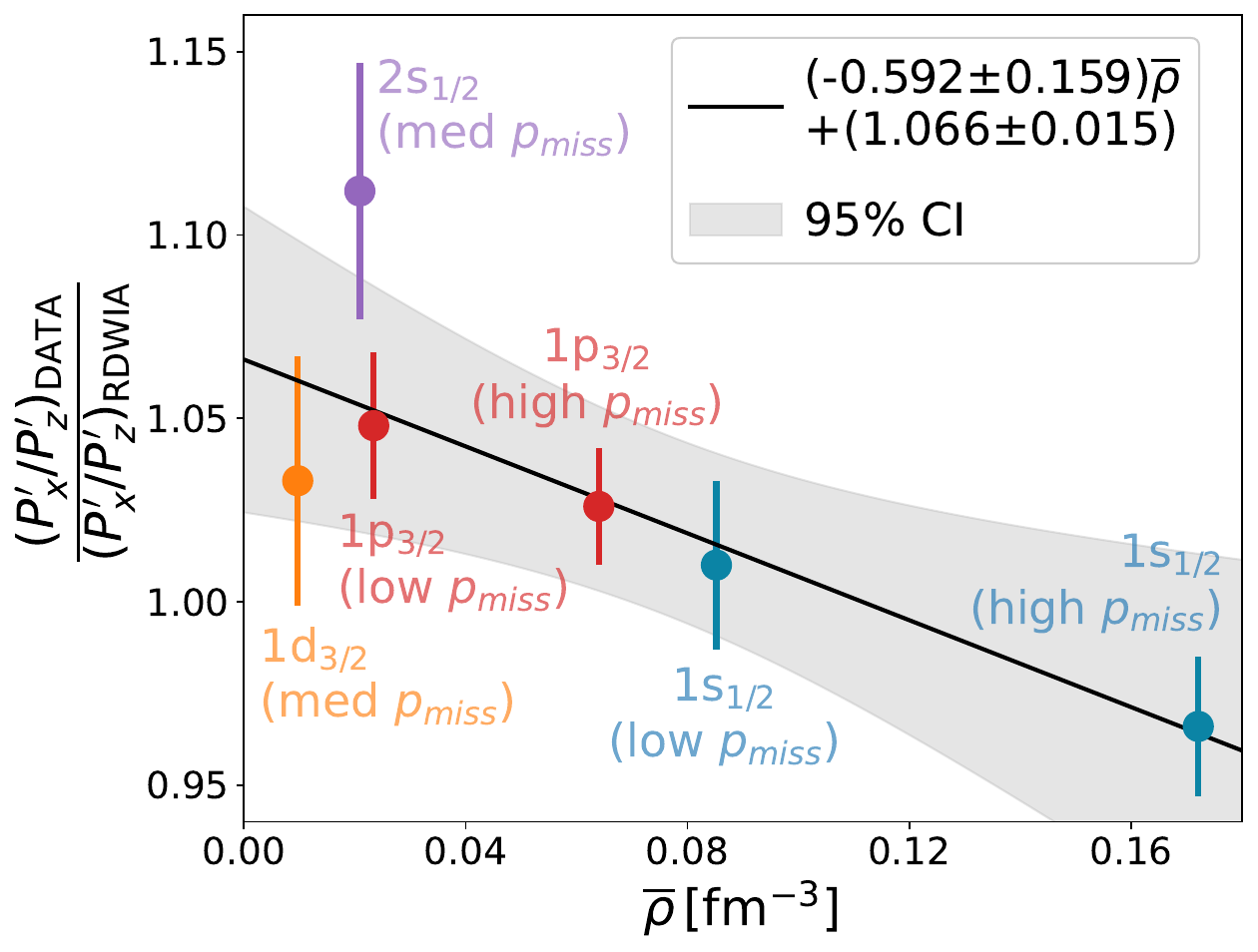}
\caption{Single shell double ratios between the measured polarization transfer components and the ones calculated with the RDWIA model from~\cite{Meucci:2001qc} as a function of the effective local nuclear density. From the fitted linear function and its $95\%$ confidence level band, a clear signature of a density-dependent effect with a negative slope can be seen. }
\label{fig:single_histogram}
\end{figure}

In parallel and anti-parallel quasi-elastic kinematics and in the one-photon exchange approximation, the calculated $P^{\prime}_x/P^{\prime}_z$ ratios depend linearly on $G_E/G_M$, which is the case for the data used in this work. The single-shell double ratios between the measured polarization transfer and those calculated in RDWIA with free-proton form factors, $(P^{\prime}_x/P^{\prime}_z)_{\rm Data}/(P^{\prime}_x/P^{\prime}_z)_{\rm RDWIA}$, are shown in Fig.~\ref{fig:single_histogram} as a function of the effective local nuclear density (Fig.~\ref{fig:pmiss_and_density}). These values were obtained as a weighted average over several $p_{\rm miss}$ bins for a better comparison between experimental and theoretical results over the kinematic phase space.
\begin{figure*}[t]
\includegraphics[width=\textwidth]{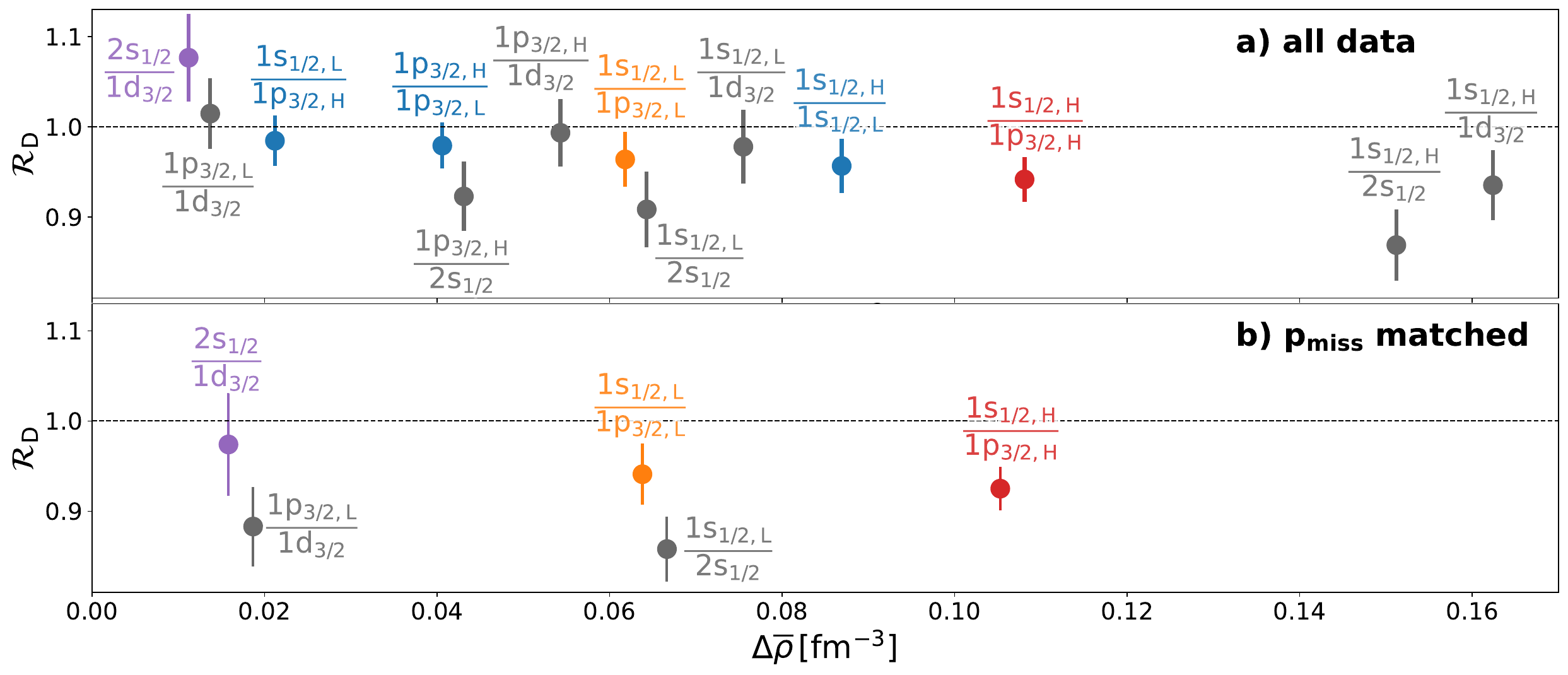}
\caption{The super ratio from Eq.~(\ref{eq:superratioD}) as a function of density difference between the two shells. The ratios are constructed with denser shell being always in the numerator of measured and calculated double ratios. Ratios between shells of the same nucleus are shown in color, while those formed for shells across the two nuclei are shown in gray. Subscripts H and L next to $^{12}\mathrm{C}$ shells denote high- and low-$p_{\rm miss}$ settings, respectively.
}
\label{fig:double_histogram}
\end{figure*}
The double-ratio results indicate a statistically-significant linear decrease as a function of the effective nuclear density with a slope of ($-0.59\pm0.16)\,\mathrm{fm}^3$. At $\overline{\rho}=0$, the fit has a value of 
$1.066\pm0.015$ where it is expected to be unity. This can be explained by the $2\%$ systematic uncertainties of the data~\cite{ceepComp,ceepTim,caeepTim}, and possible deficiencies in the models which may systematically underestimate the data. We check the consistency of these data with a proposed reduction of the effective $G_E/G_M$ ratio~\cite{Ron:2012cp} in the last section of this paper. Nevertheless, discrepancies at the level of a few percent are not necessarily a consequence of in-medium modifications. We cannot rule out RDWIA deficiencies (non-density dependent) that might contribute to the observed slope. To reduce effects of such possible deficiencies we study also the super ratios presented bellow.

%----------------  SUBSECTION ----------------
\subsection{Two-shell comparison with RDWIA}
Kinematic variations may affect the double ratio between two shells, A and B, $(P^{\prime}_x/P^{\prime}_z)_{\rm A}/(P^{\prime}_x/P^{\prime}_z)_{\rm B}$. In addition, many-body effects like FSI may be different for different shells. Those considered in the model can be largely factored out by dividing the experimental double ratio by the calculated one. Furthermore, any theoretical discrepancies common to various shells would also cancel. Thus, the \textit{super ratio}, 
\begin{equation}
\label{eq:superratioD}
    \mathcal{R}_{\rm D}=\dfrac{\left[(P^{\prime}_x/P^{\prime}_z)_{\rm A}/(P^{\prime}_x/P^{\prime}_z)_{\rm B}\right]_{\rm Data}}{\left[(P^{\prime}_x/P^{\prime}_z)_{\rm A}/(P^{\prime}_x/P^{\prime}_z)_{\rm B}\right]_{\rm RDWIA}}\,,
\end{equation}
is expected to have an improved sensitivity to the bound proton properties over other many-body effects. It allows a better comparison between a free and a bound proton. Because the calculations are performed using the elastic proton FFs, the super ratio provides a measure of the relative deviation of the effective FF ratio in the two shells.

Since medium modifications are expected to be small, at a few percent level~\cite{Ron:2012cp}, a measurement in a single configuration may not have sufficient statistical precision to observe such effects. However, the various measurements carried over different regions of $p_{\rm miss}$ and nuclei, probing different effective local nuclear densities, allow us to do a systematic study of the super ratios. Each measurement is characterized by different effective nuclear densities (see Fig.~\ref{fig:pmiss_and_density}). In Fig.~\ref{fig:double_histogram}a we show super ratios comparison of higher- to lower-density shells as a function of the \textit{difference} in the probed density. The \textit{super ratios} are shown for two measurements on $1s_{1/2}$ and $1p_{3/2}$ protons in $^{12}\mathrm{C}$, covering low- and high-$p_{\rm miss}$ ranges, and for a measurement of $2s_{1/2}$ and $1d_{3/2}$ protons in $^{40}{\rm Ca}$ (see Table~\ref{tab:kinematics}).

As inferred from the negative slope of the linear density dependence of the ratios shown in Fig.~\ref{fig:single_histogram}, the super ratios are on average below unity and tend to decrease with increasing density difference. We observe in Fig.~\ref{fig:pmiss_and_density} that between the $1s_{1/2}$ and $1p_{3/2}$ shells in $^{12}\mathrm{C}$ the density differs progressively with increase of $p_{\rm miss}$. The densities of the $2s_{1/2}$ and $1d_{3/2}$ shells in $^{40}\mathrm{Ca}$ stay much closer over the measured $p_{\rm miss}$ range. Accordingly, we observed deviations in the ratios between $^{12}\mathrm{C}$ shells being larger at high $p_{\rm miss}$ than at low $p_{\rm miss}$ and almost no effect present in the ratio between the $s$ and $d$ shells in $^{40}\mathrm{Ca}$ where densities are comparable. This suggests the presence of a density-dependent effect not yet included in RDWIA calculations.

\begin{figure*}
\includegraphics[width=\textwidth]{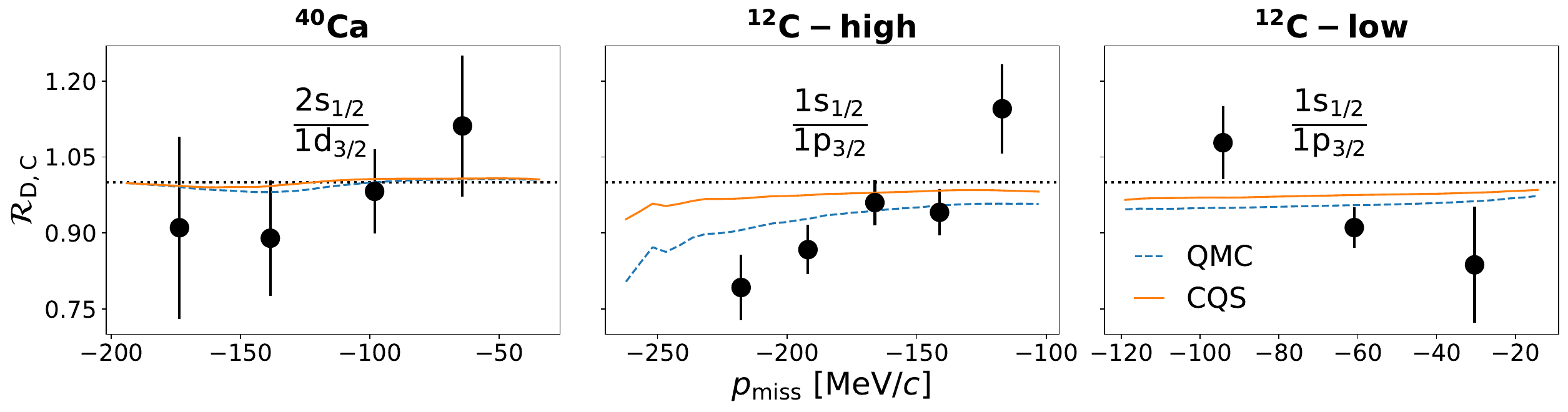}
\caption{The super ratios as a function of $p_{\rm miss}$. The data points show the super ratio from Eq.~(\ref{eq:superratioD}). The lines show the super ratios from Eq.~(\ref{eq:superratioC}) using two different models (QMC,CQS) for in-medium density-dependent modification of proton EM form factors. 
}
\label{fig:superratio_w_data}
\end{figure*}

We also present super ratios from specific shells for different nuclei; the $1s_{1/2}$ ($1p_{3/2}$) protons in $^{12}\mathrm{C}$ to those from $2s_{1/2}$ ($1d_{3/2}$) in $^{40}\mathrm{Ca}$ to probe the high (low) density difference. These \textit{cross-nuclei ratios} shown in Fig.~\ref{fig:double_histogram} are more likely to be influenced by systematic uncertainties in either measurements or calculations. Nevertheless, we see that almost all of them fall below unity and are consistent with ratios between shells of the same nucleus.

In Fig.~\ref{fig:double_histogram}b we present the same super ratios as in Fig.~\ref{fig:double_histogram}a, but we consider only data from overlapping regions of $p_{\rm miss}$ distributions of the compared shells. The overlap region was first subdivided into several bins before obtaining the super ratios and subsequently taking their weighted average. This differs from data shown in Fig.~\ref{fig:double_histogram}a, where the super ratio was formed from weighted averages of the double ratio over the entire $p_{\rm miss}$ range of each shell~\footnote{Even when considering only overlapping regions, because of varying per-bin statistical uncertainties, taking the ratio of two weighted double-ratio averages does not give the same result as a direct weighted average of the super ratio binned in $p_{\rm miss}$. This is due to the sensitivity to underlying variable changing when adding uncertainties for each bin i.e. taking whole overlap region as one bin will produce different result than if you bin it into 6 bins because of the differences in individual shell $p_{\rm miss}$ distributions}. The fact that $p_{\rm miss}$-matched super ratios are consistent with those of Fig.~\ref{fig:double_histogram}a, and the observed linear trend in Fig.~\ref{fig:single_histogram} despite the points not being ordered by $p_{\rm miss}$, further suggest that this is a density-dependent phenomenon.

%---------------------------------------------
%----------------  SECTION -------------------
%---------------------------------------------
\section{In-medium proton modifications}\label{sec:MModPredictions}
The first investigation of polarization transfer sensitivity to possible density dependent in-medium modification of nucleon form factors was carried out by Kelly~\cite{Kelly:1996hd}.This was followed by several polarization-transfer experiments.  
Measurements of the $^{4}\mathrm{He}(\vec{e},e^{\prime}\vec{p}\,)$ reaction, performed both at MAMI~\cite{Dieterich:2000mu} and at JLab~\cite{Strauch,Paolone}, favored the calculations that included medium modifications models~\cite{PhysRevLett.83.5451}. Two models were considered. The chiral quark soliton (CQS) model~\cite{Christov:1995hr,Smith:2004dn} that mainly modifies valence quark contributions and the quark-meson coupling (QMC) model~\cite{Lu:1997mu,Lu:1998tn}.  However, the same data were later described by different calculations using free-proton EM FFs~\cite{PhysRevLett.94.072303}. 

We compare our results with the calculations using QMC and CQS models within the relativistic multiple scattering Glauber approximation (RMSGA)~\cite{Ron:2012cp,Ryckebusch:2003fc}. As suggested in Ref.~\cite{Ron:2012cp}, density-dependent in-medium modifications of the bound proton should be reflected in the super ratio of shells A and B
\begin{equation}
\label{eq:superratioC}
    \mathcal{R}_{\rm C}=\dfrac{\left[(P^{\prime}_x/P^{\prime}_z)_{\rm A}/(P^{\prime}_x/P^{\prime}_z)_{\rm B}\right]_{\rm QMC,CQS}}{\left[(P^{\prime}_x/P^{\prime}_z)_{\rm A}/(P^{\prime}_x/P^{\prime}_z)_{\rm B}\right]_{\rm Free}}\,.
\end{equation}
We refer to Fig.~2 in Ref.~\cite{Ron:2012cp} and related discussion to illustrate the magnitude and density dependence of the form factor modification effects in these two models. In Fig.~3 of the same reference, it is shown how this modification is reflected in the super ratios as a function of $p_{\rm miss}$. The predictions are that the electric form factor, $G_E$, decreases with increasing nuclear density regardless of the model. This is unlike the magnetic form factor, $G_M$, which increases in the QMS model or is hardly affected within the CQS model. Combined, this results in a decrease of the form factor ratio with increasing density for both models~\cite{Ron:2012cp}. To cover all of our kinematic settings and target nuclei we extended the original RMSGA calculations. To ensure self-consistency, the effective local nuclear density experienced by the proton from each event was also obtained in RMSGA through the procedure described in Refs.~\cite{Cosyn:2009bi,Ryckebusch:2011yy} analogous to the one that was used for calculation of RDWIA densities in previous sections. 

The predicted super ratios as a function of $p_{\rm miss}$ are shown for three kinematic settings of this work in Fig.~\ref{fig:superratio_w_data}. The super-ratio, $\mathcal{R}_{\rm C}$, from Eq.~(\ref{eq:superratioC}) is calculated in RMSGA. The numerator is calculated using density-dependent EM FFs predicted by either QMC or CQS.  The denominator is obtained by using free proton EM FFs from Ref.~\cite{budd2003modeling}. We compare these predictions to our \textit{super ratios} using Eq.~(\ref{eq:superratioD}), where data are divided by our RDWIA calculations. These super-ratios are consistent with the ones predicted by the calculations using modified density-dependent form factor. The deviations from unity are more prominent in high-$p_{\rm miss}$ region of $^{12}\mathrm{C}$ shown in central panel of Fig.~\ref{fig:superratio_w_data}, where density differences between shells are the largest. The super-ratios for $^{40}\mathrm{Ca}$ are about unity as the density of the shells are about equal. This observation is in line with previous analyses~\cite{Paolone, PhysRevLett.83.5451, PhysRevC.71.014605}.

%---------------------------------------------
%----------------  SECTION -------------------
%---------------------------------------------
\section{Conclusions}
Polarization transfer to bound protons provides a sensitive tool to probe the bound proton electromagnetic form factors. However, a direct comparison of the measurements to calculations does not allow us to determine if deviations are due to many-body or in-medium modifications in the bound proton structure~\cite{ceepTim,caeepTim}. Discrepancies due to many-body effects are expected to be largely mitigated in super ratios by comparing the double ratios of polarization-transfer data to those calculated in RDWIA for individual shells in $^{12}\mathrm{C}$ ($1p_{3/2}$ and $1s_{1/2}$) and $^{40}\mathrm{Ca}$ ($1d_{3/2}$ and $2s_{1/2}$).  

In our present study of double and super ratios for shells with different effective nuclear densities we observed a systematic density-dependent deviation. We further found it to be consistent with a reduction of the EM FF ratio, $G_E/G_M$, for protons bound in higher density shells compared to those in lower density shells, as predicted in CQS amd QMC models. While we cannot fully exclude other density-dependent effects, past and future analyses of other polarization observables~\cite{InducedC12,HelictyAsymmC12Kolar} might provide more stringent limits on FSI and other many-body effects included in various calculations.

Unlike the ratio of the transverse to the longitudinal components, which is linearly dependent on the FF ratio $G_E/G_M$, the transverse and longitudinal components, individually, have different dependencies on the electric and magnetic FFs. Additional high statistics data of the transfer components may yield information on the individual behavior of the EM FFs of the bound proton. 

%---------------------------------------------
%----------------  SECTION -------------------
%---------------------------------------------
\begin{acknowledgments}
%\section*{Acknowledgements}
We would like to thank the Mainz Microtron operators and the technical crew for the excellent operation of the accelerator. This work is supported by the Israel Science Foundation (Grant 951/19) of the Israel Academy of Arts and Sciences, by the Israel Ministry of Science, Technology and Spaces, by the PAZY Foundation (Grant No. 294/18), by the Deutsche Forschungsgemeinschaft (Collaborative Research Center 1044), by the U.S. National Science Foundation (Grants No. PHY-2111050, No. PHY-2111442, and No. PHY-2239274), and by the  United States-Israeli Binational Science Foundation (BS) as part of the joint program with the NSF (Grant No. 2020742). We acknowledge the financial support from the Slovenian Research Agency (research core funding No.~P1-0102).  
\end{acknowledgments}
\FloatBarrier

%\section{References}
%\newpage{}
%\bibliographystyle{elsarticle-num}
%\small \addcontentsline{toc}{section}{\refname}\bibliography{induced}

%\addcontentsline{toc}{section}{\refname}\small{\bibliography{hdep}}
%\clearpage
\bibliography{hdep}

\end{document}